\documentclass[twocolumn,showpacs,floatfix,superscriptaddress,amsmath,amssymb]{revtex4}
\usepackage{amsmath}
\usepackage{amsfonts}
\usepackage{amssymb}
\usepackage{graphicx}
\begin{document}
\title{Quantum coherence and spin nematic to nematic quantum phase transitions in biquadratic spin-$1$ and -$2$ XY chains with rhombic single-ion anisotropy}

\author{Rui Mao}
\affiliation{Centre for Modern Physics and Department of Physics,
Chongqing University, Chongqing 400044,
China}

\author{Yan-Wei Dai}
\affiliation{Centre for Modern Physics and Department of Physics,
Chongqing University, Chongqing 400044,  China}

 \author{Sam Young Cho}
 \email{sycho@cqu.edu.cn}
\affiliation{Centre for Modern Physics and Department of Physics,
Chongqing University, Chongqing 400044,
China}

 \author{Huan-Qiang Zhou}
\affiliation{Centre for Modern Physics and Department of Physics,
Chongqing University, Chongqing 400044,
China}

\begin{abstract}
 We investigate quantum phase transitions and quantum coherence in infinite biquadratic spin-$1$ 
 and -$2$ XY chains with rhombic single-ion anisotropy.
 All considered coherence measures such as the $l_1$ norm of coherence, the relative entropy of coherence, and the quantum Jensen-Shannon divergence, and the quantum mutual information show consistently that singular behaviors occur for the spin-$1$ system, which enables to identity quantum phase transitions.
 For the spin-$2$ system, the relative entropy of coherence and the quantum mutual information properly detect no singular behavior in the whole system parameter range, while the $l_1$ norm of coherence
 and the quantum Jensen-Shannon divergence show a conflicting singular behavior of their first-order derivatives.
 Examining local magnetic moments and spin quadrupole moments lead to the explicit identification of novel orderings of spin quadrupole moments with zero magnetic moments in the whole parameter space.
 We find the three uniaxial spin nematic quadrupole phases for the spin-$1$ system and the two biaxial
 spin nematic phases for the spin-$2$ system.
 For the spin-$2$ system,
 the two orthogonal biaxial spin nematic states are connected adiabatically without an explicit phase transition, which can be called quantum crossover. The quantum crossover region is estimated by using
 the quantum fidelity.
 Whereas
 for the spin-$1$ system, the two discontinuous quantum phase transitions occur
 between three distinct uniaxial spin nematic phases.
 We discuss the quantum coherence measures and the quantum mutual information
 in connection with the quantum phase transitions including the quantum crossover.

\end{abstract}


\maketitle

\section{Introduction}
  Quantum phase transitions \cite{Sachdev} lie at the heart of
  quantum many-body phenomena in condensed matter physics.
  In contrast to classical phase transitions induced by thermal fluctuations,
  quantum fluctuations
  originated from the Heisenberg uncertainty principle give rise to
  quantum phase transitions for varying system parameters
  at zero temperature.
 Especially, the investigations of various quantum spin systems
 have revealed many novel states and quantum phase transitions for quantum magnetism \cite{Book_Schollwock}.
 Of particular interest are quantum nonmagnetic phases of matter
 such as, for instance, Haldane phase \cite{Haldane,Pollmann,Tonegawa,Okamoto,Tzeng,Kjall},
 dimerized phase \cite{Barber,Klumper,Chubukov,Solyom},
 and spin nematic (or quadrupolar) phase \cite{Blume,Haramoto,Chen71,Nakatsuji,Podolsky,Tsunetsugu,Bhattacharjee,Lauchli,Michaud,
 Harada,Toth,Gong,Lee,Niesen,Fridman}
 because
 quantum magnetism normally originates from the spin exchange
 coupling between quantum spins
 but such exotic states have no conventional long-range magnetic order, i.e.,
 spin correlations are short-ranged.
 Furthermore, in manipulating cold atoms in optical lattice, the recent impressive progresses \cite{Stenger,Stamper,Chang,Griesmaier,Greiner} have made various quantum spin nematics  \cite{Song,Demler,F_Zhou,Barnett,Bernier,Tu,Turner}
 realizable among other exotic states experimentally.

 Essentially, quantum phase transitions are abrupt
 changes of groundstate wavefunction structure driven by quantum fluctuations.
 Such changes of groundstate wavefunction structure lead to similar sudden changes of correlations.
 For example, in the groundstate of the transverse-field Ising model  at zero temperature,
 the spin-spin correlation is long-ranged, indicating the spin ordering, but this disappears exponentially when the transverse field exceeds the critical
 value \cite{Pfeuty}.
 Though a few correlations are known involving for quantum phase transitions in quantum many-body systems such as in the transverse-field Ising model, general quantum many-body systems have various correlations of very different nature that result from interactions and thus finding proper correlation related to phase transition is to characterize quantum phase transition.
 In this sense, from the information theory perspective, for instance,
 quantum mutual information measuring the sum of quantum and classical correlations can be used as a general tool to identify quantum phase transitions
 because it is not required to know a priori what the right correlation function is
 in a given many-body system \cite{Groisman,Adami,Alcaraz,Schumacher,Dai1,Dai2}.

  Recent quantification of quantum coherence \cite{Baumgratz} in quantum information
  science has also lead to disclose intriguing connections between quantum coherence
  and correlation \cite{Ma,Streltsov,Xi,Jeong}.
  Another aspect of fundamental feature of quantum phases and quantum phase transition in a many-body system can be thus explored
 from a perspective of quantum coherence as a fundamental property of quantum mechanics.
 Moreover, quantum coherence exhibits even in separable product states \cite{You18} without quantum entanglement (correlation) \cite{Wootters},
 which implies that it can see a different feature of quantum phases and quantum phase transitions.
 In fact, a variety of quantum coherence measures are introduced including
 the $l_1$ norm of coherence \cite{Baumgratz}, the relative entropy of coherence \cite{Baumgratz}, and
 the quantum Jensen-Shannon divergence \cite{Radhakrishnan}.
 Such various quantum coherence measures have been studied in detecting and characterizing quantum
 phase transitions for several spin chain models such as
 the transverse-field Ising model \cite{JJChen},
 the spin-$1/2$ anisotropic XY chain \cite{Karpat}, the two-dimensional Kitaev
 honeycomb model \cite{QChen},
 the anisotropic spin-$1/2$ Heisenberg XYZ chain with the Dzyaloshinskii-Moriya (DM) interaction in magnetic fields \cite{Yi,Thakur},
 the spin-$1/2$ XY chain with DM interaction under magnetic fields \cite{Radhakrishnan},
 the spin-$1/2$ XY model with three-spin interaction \cite{Sha,Li}
 and a transverse magnetic
 field \cite{Hu},
 the compass chain under an alternating magnetic field \cite{You18}, and
 the spin-$1$ XXZ chain \cite{Malvezzi,Chao} and bilinear-biquadratic chain \cite{Malvezzi}.
 Accordingly, as an example, the continuous (or second-order) quantum phase transition belonging to
 the Ising universality class in the spin-$1/2$ XY model
 has shown to be captured by using quantum coherence measures such as
 the derivative of quantum coherence quantified by
 the $l_1$ norm of coherence \cite{Sha}, the relative entropy \cite{JJChen},
 the quantum Jensen-Shannon divergence \cite{Radhakrishnan},
  and
 the skew information \cite{Karpat}.

 In fact, many of such studies on quantum coherence in quantum spin systems
 have been performed for spin-$1/2$ systems based on available analytical solutions.
 Higher spin systems have then received relatively less attentions on exploring connections between quantum
 coherence and quantum phase transitions.
 In contrast with spin-$1/2$ systems, however
 it is known that the presence of biquadratic interaction can induce a quadrupole order for spins higher than $1/2$ \cite{Blume}.
 For example, the Heisenberg model with biquadratic interaction
 possesses a quadrupole phase in two-dimensional spin lattice systems for spin-$1$
 \cite{Tsunetsugu,Toth,Gong,Lee,Niesen} and for spin-$3/2$ \cite{Fridman},
 and in three-dimensional spin lattices for spin-$1$ \cite{Harada}.
 Thus to understand various aspect of fundamental features of quantum phase transitions,
 the purpose of this paper is to investigate quantum coherence and
 quantum phase transition in spin nematic states without the conventional long-range magnetic order.

 In our study, we then introduce one-dimensional infinite spin-$1$ and spin-$2$ biquadratic XY models
 with rhombic single-ion anisotropy.
 In order to numerically investigate these infinite-lattice systems,
 we employ the infinite matrix product state (iMPS) representation  with
 the infinite time-evolving block decimation (iTEBD) method \cite{Vidal03,Vidal07,Su12}.
 Our numerical results show that the spin-$1$ chain and the spin-$2$ chain reveal fundamental difference
 of their phase diagrams each other.
 Connections between quantum phase transitions and
 tools of quantum information theory are studied for our spin models by calculating
 the $l_1$ norm of coherence, the relative entropy of coherence,
 the quantum Jensen-Shannon divergence, and the quantum mutual information.
 The quantum coherence measures show
 their characteristic features that
 are the two discontinuities for the spin-$1$ system
 and the incoherent point of zero coherence for the spin-$2$ system.
 However, in their first-order derivatives at the incoherent point for
 the spin-$2$ system,
 the relative entropy of coherence exhibits a non-singular behavior,
 while the $l_1$ norm of coherence and the quantum Jensen-Shannon divergence
 have a singular behavior conflictingly.
 Supportively to the relative entropy of coherence,
 the quantum mutual information show the two discontinuous jumps for the spin-$1$ system
 and the monotonous hill-shape without any singular behavior for the spin-$2$ system.
 Also, the groundstate energy per site reveals
 the characteristic features indicating a first-order quantum phase transition for the spin-$1$ system.
 For the spin-$2$ system,
 the frist- and the second-order derivatives of the groundstate energy per site do not show
 any non-analytical feature of the groundstate energy per site indicating
 quantum phase transition at the corresponding zero coherence point.
 We find that for the spin-$1$ system, the local quadrupole order parameters
 reveal the three quadrupolar ordered phases and exhibit the discontinuous quantum phase transitions between
 the three distinct uniaxial spin nematic phases in agreement with the results of the tools
 of quantum information theory.
 While for the spin-$2$ system, the two biaxial spin nematic states are connected adiabatically without explicit phase transition at a specific parameter
 in spite of the orthogonal nematic states, which can be called quantum crossover.
 The quantum coherence measures and the quantum mutual information are discussed in connection with the
 quantum crossover.

 This paper is organized as follows.
 In Sec. \ref{section2}, the infinite biquadratic XY chains with rhombic single-ion anisotropy
 is introduced for spin $1$ and $2$.
 A brief explanation of the iMPS approach is given
 in calculating groundstate wavefunctions for the infinite chain models.
 Section \ref{section3} devotes to discuss the behaviors of the quantum coherence measures and
 the quantum mutual information.
 To clarify the relationship between quantum phase transitions and quantum coherence measures,
 we consider the groundstate energy per site and the bipartite entanglement entropy
 in our iMPS approach in Sec. \ref{section4}.
 In Sec. \ref{section5}, the local magnetization and
 the local quadrupole order parameters are discussed to clarify
 the uniaxial and biaxial spin nematic
 phases and the quantum phase transitions in association with the quantum coherence measures and
 the quantum mutual information.
 The very different features between the spin-$1$ and -$2$ systems
 are given by discussing the detailed behaviors of the uniaxial and biaxial spin nematic states.
 A summary and remarks of this work is given in Sec. \ref{summary}.

\section{Biquadratic XY chains and iMPS approach}
\label{section2}

 We start with the one-dimensional biquadratic spin XY models with rhombic single-ion anisotropy.
 The Hamiltonian can be written as
\begin{equation}
 H = -J\sum_{i=-\infty}^\infty \left(S^{x}_iS^{x}_{i+1}+ S^{y}_{i}S^{y}_{i+1}\right)^2
      + R \sum_{i=-\infty}^\infty [ (S^x_i)^2 - (S^y_i)^2 ],
\label{Ham}
\end{equation}
 where $J (>0)$ is the biquadratic exchange interaction and $S^\alpha_i$ is the spin-$1$ or -$2$ operator on the $i$th site for $\alpha \in \{x,y,z\}$.
 The rhombic single-ion anisotropy is denoted by $R$, which is normally referred to as zero-field splitting parameter due to a crystal-field anisotropy.
 The rhombic single-ion anisotropy effect has been very recently investigated in the
 spin-$1$ Heisenberg model \cite{Tzeng17} and XXZ model \cite{Ren}.
 We will study the same form of the Hamiltonian (\ref{Ham})
 for the spin-$1$ and-$2$ systems, respectively.

 For our one-dimensional infinite lattices of the spin chains,
 a wave function $\left|\psi\right\rangle$ of the Hamiltonian
 can be represented in the iMPS.
 By employing the iTEBD method,
 a numerical groundstate $\left|\psi_G\right\rangle$ can be obtained in the iMPS representation
 \cite{Su12,Su13,Wang,Dai17}.
 When the initially chosen state approaches to a groundstate,
 the time step is chosen to decrease from $dt = 0.1$ to $dt = 10^{-6}$
 according to a power law.
 Once the system energy converges to a groundstate energy, which yields a groundstate wavefunction in the iMPS representation
 for a given truncation dimension, i.e., here $\chi=30$.
 The iMPS groundstate wave function $|\psi_G\rangle$ is the full description of the groundstate in a pure state.
 The full density matrix $\varrho_G = |\psi_G\rangle\langle\psi_G|$
 gives any reduced density matrix $\varrho_{L}$ for lattice-block $L$
 by tracing out the degrees of freedom of the rest of the
 lattice-block $L$, i.e., $\varrho_{L} = \mathrm{Tr}_{L^c} \, \varrho_G$.
 In our study, single-site reduced density matrix is used because
 the reduced density matrices for lattice blocks give no significant changes
 of the values of quantum coherence measures compared with the single-site
 coherence measures and
 a single-spin coherence can be experimentally accessible without
 requirements of full tomography of the state.
 Two-site reduced density matrix is used for quantum mutual information.

%
\begin{figure}
\includegraphics [width=0.45\textwidth]{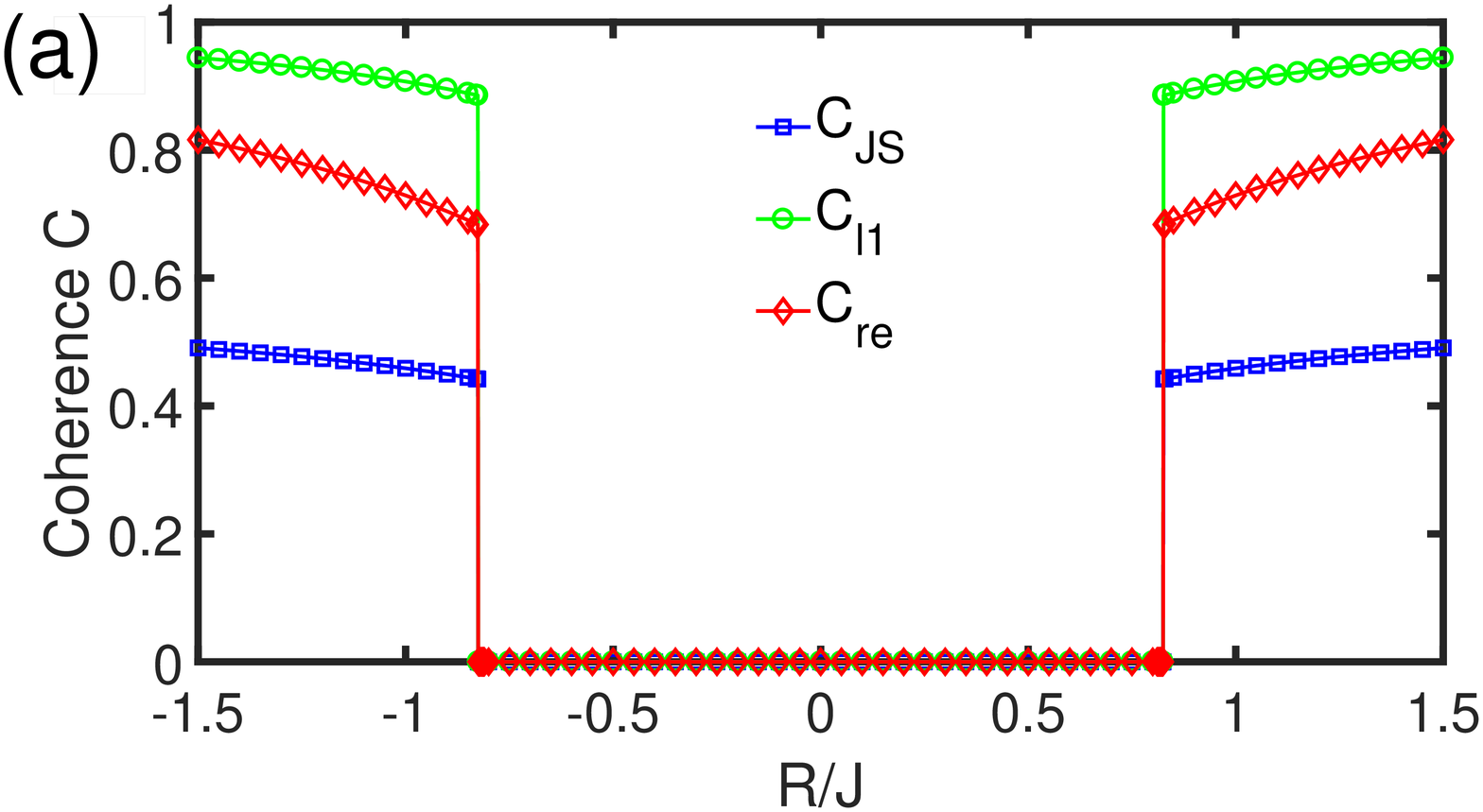}
\includegraphics [width=0.45\textwidth]{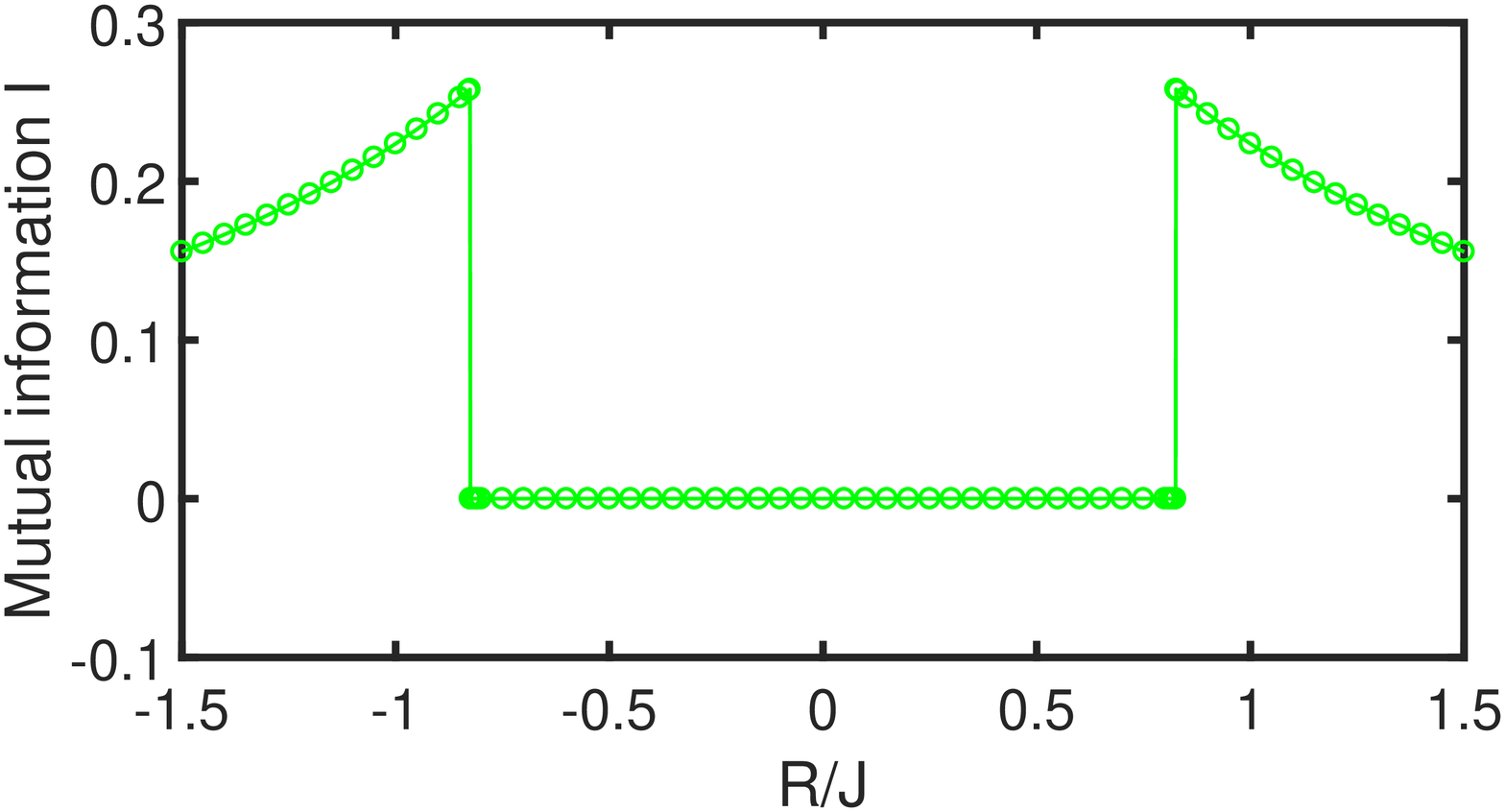}
\includegraphics [width=0.45\textwidth]{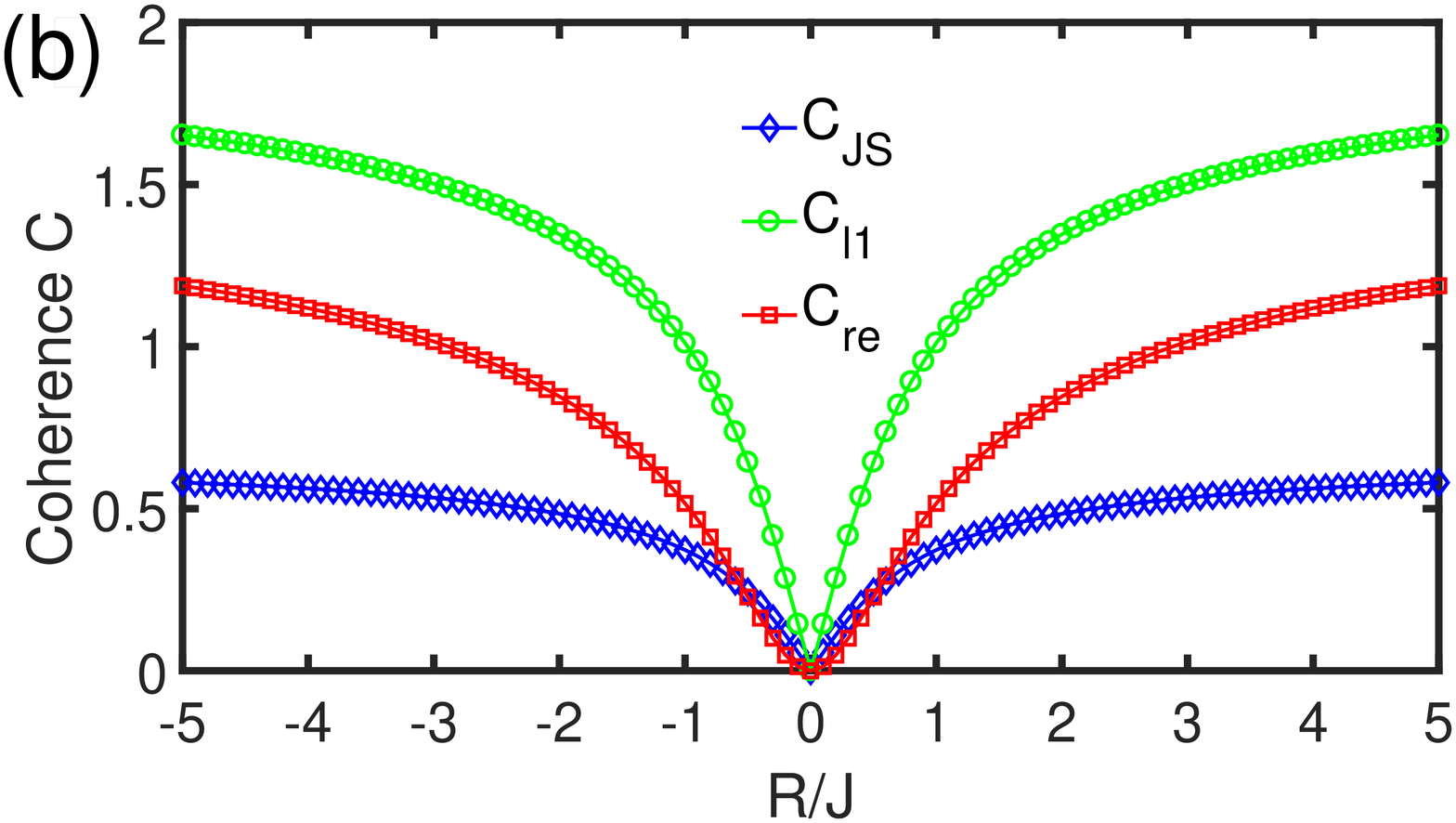}
\includegraphics [width=0.45\textwidth]{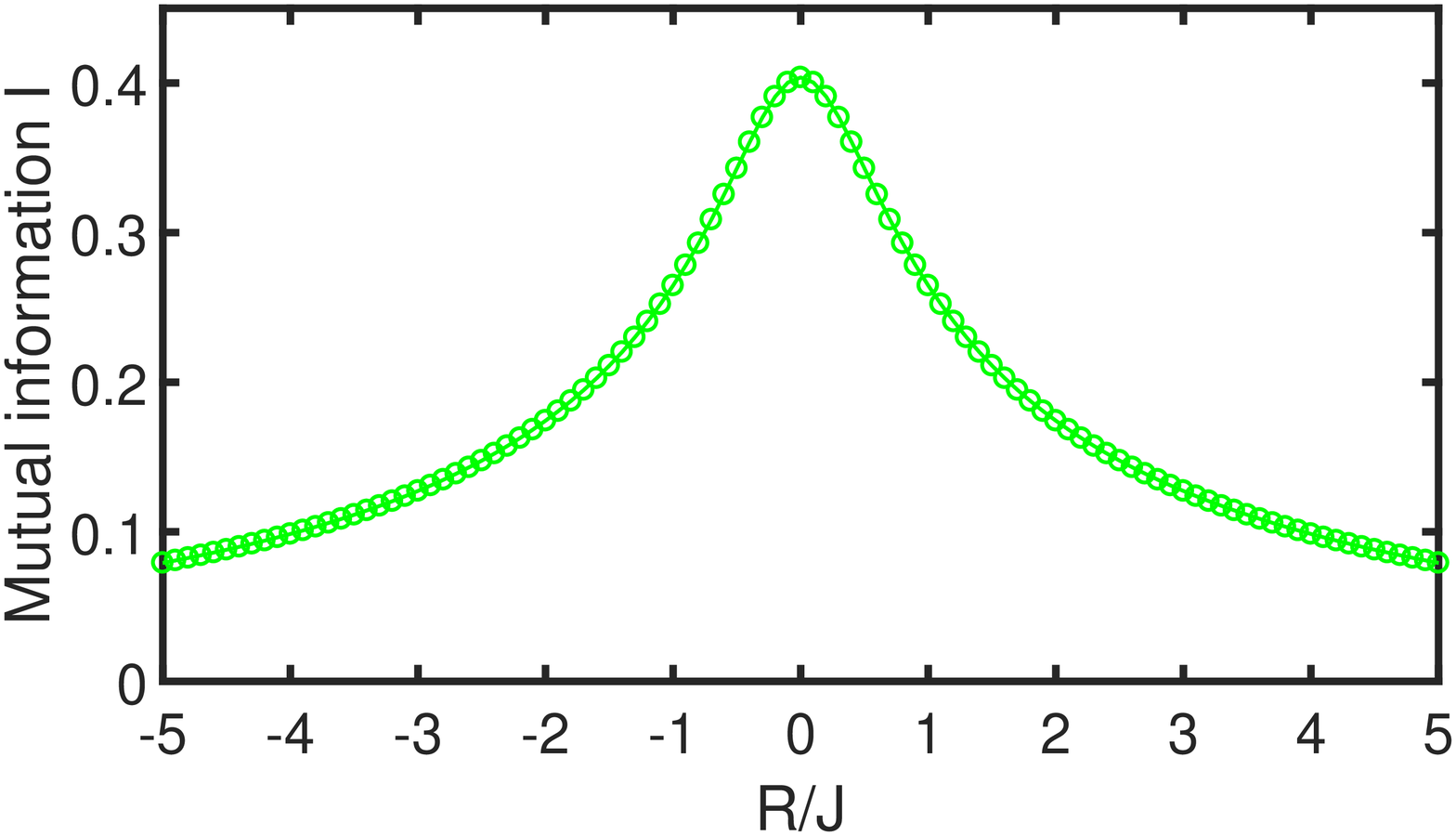}
\caption{ (Color online)
 Quantum coherence measures and quantum mutual information as
 a function of $R/J$ for (a) the spin-$1$ system and (b) the spin-$2$ system.
}
 \label{fig1}
\end{figure}

\section{Quantum coherence measures, quantum mutual information and quantum phase transition}
\label{section3}
 Over the last two decades, quantum information science has developed to embody quantum physical phenomena  in quantum information resources that can be exploited to achieve tasks that are beyond the realms of classical physics.
 Such developments have led to provide quantitative measures for quantum coherence and correlation.
 The qualitative measures have been implemented to explore another aspects of nature of critical phenomena
 in quantum many-body systems.
 In particular, quantum coherence measures and quantum mutual information
 have been investigated whether they can capture quantum phase transition and its criticality.
 In addition, they has been studied to characterize
 a nontrivial groundstate such as factorized state or product state.
 Various quantum coherence measures,
 including the relative entropy of coherence, the $l_1$ norm quantum coherence,
 and the Jensen-Shannon divergence,
 are suggested in different notions of incoherent operations.
 In our spin chain models, we investigate such quantum coherence measures provided in the alternative
 frameworks for characterizing quantum phase transitions.

\subsection{Quantum coherence measures}
 Once a reduced density matrix $\varrho$ is obtained from iMPS groundstate wavefunctions,
 one can calculate quantum coherence measures based on the reduced density matrix.
 For comparison, we consider the three coherence measures, i.e.,
  the $l_1$ norm of coherence $C_{i_1}(\varrho)$ \cite{Baumgratz}, the relative entropy of coherence $C_{re}(\varrho)$ \cite{Baumgratz}, and the quantum Jensen-Shannon divergence $C_{JS}(\varrho)$ \cite{Radhakrishnan}.
 As a geometric measure that can be used as a formal distance measure,
 the $l_1$ norm of coherence $C_{l_1}(\varrho)$ is given as the sum of absolute values of all off-diagonal elements of the density matrix $\varrho$, i.e.,

\begin{equation}
 C_{l_1}(\varrho) = \sum_{n \neq m} |\varrho_{n m}|.
\end{equation}
For a given basis, as a valid measure of coherence,
the relative entropy of coherence $C_{re}(\varrho)$ is given by
\begin{equation}
 C_{re}(\varrho) = S(\varrho\, \| \,\varrho_{diag})= S(\varrho_{diag}) - S(\varrho),
\end{equation}
 where removing all off-diagonal entries of $\varrho$ gives the incoherent state $\varrho_{diag}$ corresponding to the state $\varrho$.
 $S(\varrho)= - \mathrm{Tr} \varrho \log_2 \varrho$
 is the von Neumann entropy.
 Together with these two coherence measures, we also consider the quantum Jensen-Shannon divergence given as
\begin{equation}
 C_{JS}(\varrho) = \sqrt{S\left(\frac{\varrho_{diag}+\varrho}{2} \right)- \frac{S(\varrho_{diag})+S(\varrho)}{2}}.
\end{equation}

\begin{figure}
\includegraphics [width=0.45\textwidth]{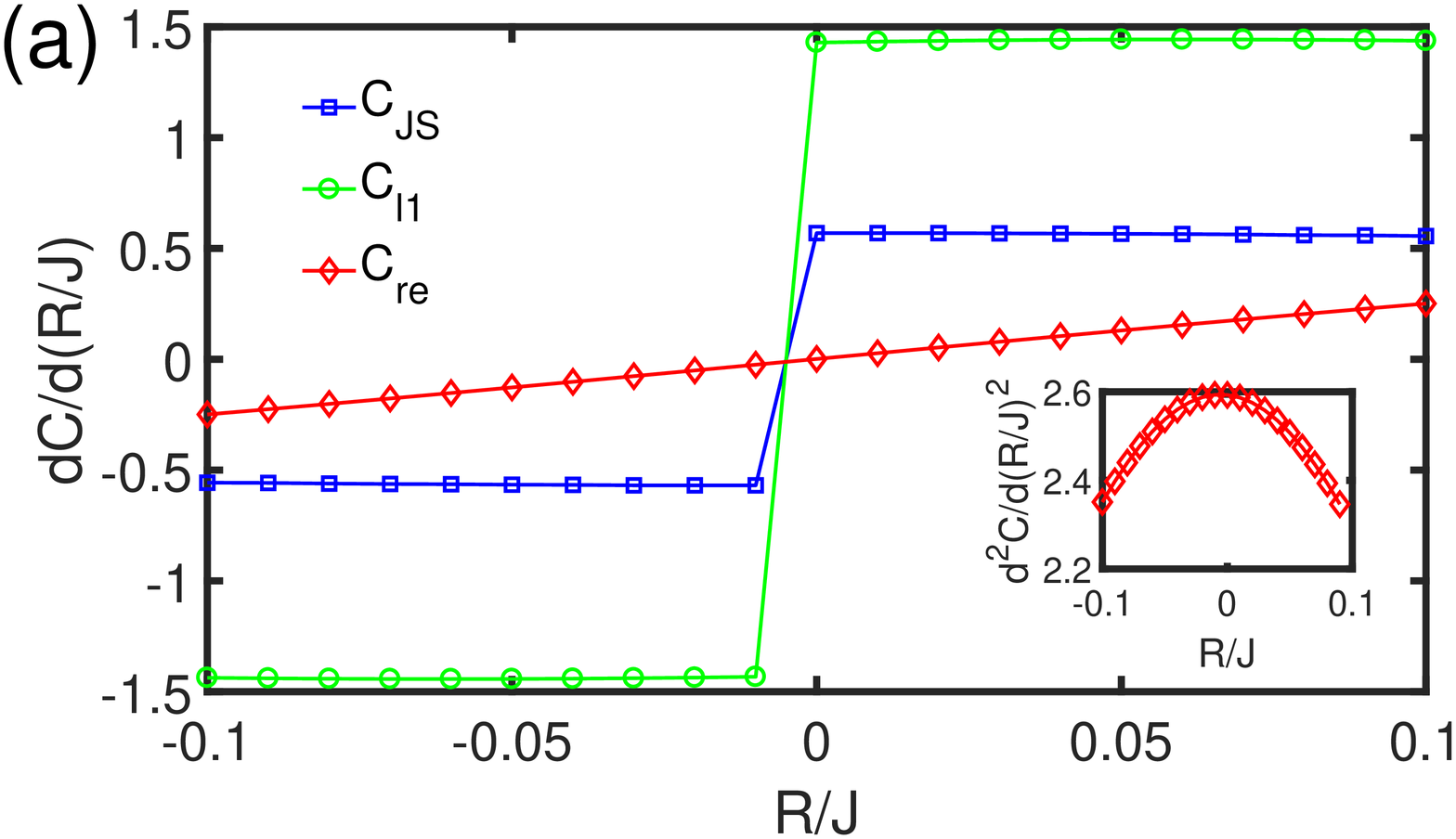}
\includegraphics [width=0.45\textwidth]{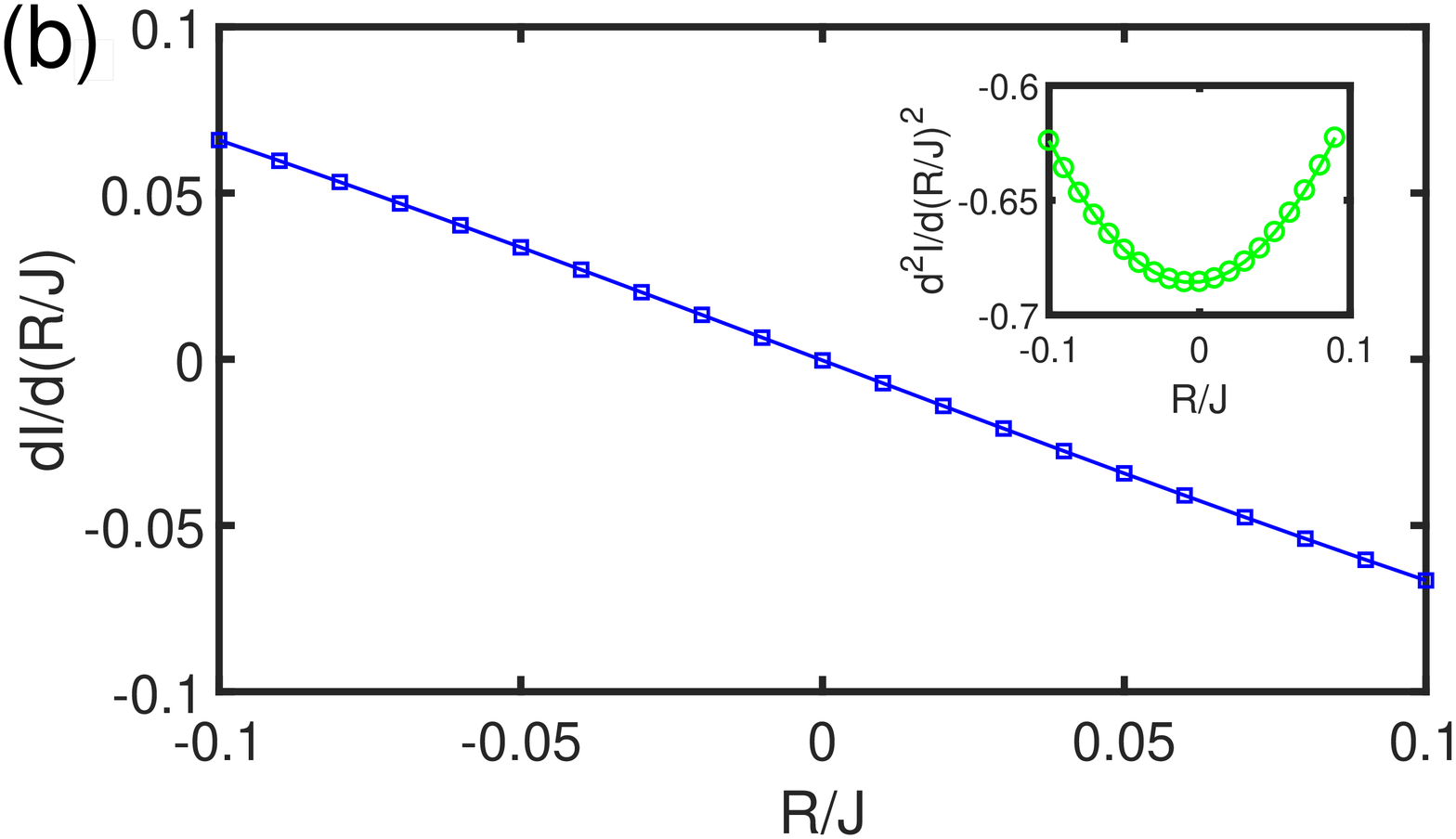}
\caption{ (Color online)
 First-order derivative of (a) three quantum coherence measures $C$
 and (b) quantum mutual information $I$ as a function of $R/J$ for the spin-$2$ system.
 In the insets of (a) and (b), the second-order derivatives of the relative entropy of coherence and
 the quantum mutual information are plotted, respectively.
}
 \label{fig2}
\end{figure}
 We plot the three coherence measures as a function of $R/J$
 for the biquadratic XY chains with rhombic single-ion anisotropy in Fig \ref{fig1}.
 For the spin-$1$ system, Fig. \ref{fig1}(a) shows that all three coherence measures undergo
 two abrupt jumps at $R = \pm 0.826 J \equiv \pm R_c$.
 These non-analytical points indicate that discontinuous phase transitions occur at the two points. Furthermore, three coherence measures become zero simultaneously, i.e.,
 $C_{l_1}=C_{re}=C_{JS}=0$ for $ -R_c < R < R_c$.
 Then the groundstate is in an incoherent state for $ -R_c < R < R_c$.

 In contrast to the case of the spin-$1$ system,
 however Fig. \ref{fig1}(b) shows that the three coherence measures exhibits no such abrupt jumps
 for the spin-$2$ system.
 At $R/J=0$, the three coherence measures are zero simultaneously
 and thus the groundstate is in an incoherence state.
 Only the noticeable inflexion of the three coherence measures occurs at $R/J=0$.
 In order to confirm whether the incoherent point $R/J=0$ relates to a quantum phase transition,
 we perform the numerical derivatives of the three coherence measures.
 As shown in Fig. \ref{fig2}(a),
 the first-order derivatives of the $C_{l_1}$ and $C_{JS}$ show
 their non-analytical behaviors at $R/J=0$, which may indicate
 an occurrence of quantum phase transition, while the relative entropy of coherence $C_{re}$
 shows a monotonous change across the incoherent point.
 Even the second-order derivative of the $C_{re}$ does not reveal any radical change
 indicating the non-analyticity of $C_{re}$,
 as shown in the inset of Fig. \ref{fig2}(a).

 At this stage, for the spin-$2$ system, it cannot be determined whether
 any quantum phase transition occurs or not because the non-analyticity of the $C_{re}$
 cannot be determined at $R/J=0$ up to the second-order derivative of the $C_{re}$.
 The other reason is why simply we cannot rule out a possibility of
 occurring a higher-order quantum phase transition such as,
 for instance,
 a Berezinskii-Kosterlitz-Thouless (BKT)-type quantum phase transition
 known as an infinite-order quantum phase transition
 \cite{You18,Radhakrishnan,Li}
 because actually the non-analyticity of
 the quantum information theoretic quantities defined by the reduced density matrices
 is connected to the non-analyticity of the groundstate energies
 of quantum many-body systems through the reduced density matrix and its derivatives \cite{Wu}.
 Thus from our results at $R/J=0$, it is hard to determine whether the
 $C_{re}$ is non-analytic or not
 because numerically reaching to a reliable much higher derivative
 is to be a very difficult task.
 This issue will then be clarified in the following sections.

 Finally, for the both spin-$1$ and-$2$ systems in Fig. \ref{fig1}, one can notice that
 the $l_1$ norm of coherence $C_{l_1} (\varrho)$ is bigger than the relative entropy of coherence $C_{re}(\varrho)$ except for that all coherence measures are zero in
 the range of $ -R_c < R < R_c$ for the spin-$1$ system and at the incoherent point $R/J=0$
 for the spin-$2$ system. This implies that $C_{l_1} (\varrho) \geq C_{re}(\varrho)$ holds, which was conjectured and proved only for the pure states and qubit states in Ref. \cite{Rana}. For a mixed state, the validity of the conjecture was shown in a compass chain under an alternating magnetic field in Ref. \cite{You18}.
 Thus our result support that the
 $l_1$ norm of coherence $C_{l_1} (\varrho)$ is an upper bound for
 the relative entropy of coherence $C_{re}(\varrho)$, i.e.,
 $C_{l_1} (\varrho) \geq C_{re}(\varrho)$ for our mixed states.

\subsection{Quantum mutual information}
 In the previous subsection, we have studied the quantum coherence measures
 in detecting quantum phase transitions.
 Interestingly, the spin-$2$ system exhibits an inconsistent behavior on
 the quantum coherence measures.
 At the incoherent point $R/J=0$,
 the non-analytical behaviors of the $l_1$ norm of coherence \cite{Sha} and the quantum Jensen-Shannon divergence \cite{Radhakrishnan} can be interpreted as an occurrence of quantum phase transition.
 However, the non-analyticity of the $C_{re}$ could not be demonstrated.
 As we discussed in the introduction,
 correlations can also undergo an abrupt change for quantum phase transitions.
 Thus in this subsection,
 we will consider a generalized correlation, i.e., the quantum mutual information that
 is based on entanglement entropy and measures a total sum of classical and quantum correlations without knowing a proper correlation operator.
 Using the von Neumann entropy, the quantum mutual information between two sites $A$ and $B$
 can be defined as
\begin{equation}
 I(A:B) = S_{A} + S_{B} - S_{AB},
 \label{Mutual}
\end{equation}
 where $S_{A/A\cup B} =- \mathrm{Tr} \varrho_{A/A\cup B} \log_2 \varrho_{A/A\cup B}$
 are the von Neumann entropies with the reduced density matrix $\varrho_{A/A\cup B}$ for one site $A$ and two sites $A\cup B$, respectively. This quantum mutual information can be used to detect and characterize
 quantum phase transitions \cite{Groisman,Adami,Alcaraz,Schumacher,Dai1,Dai2}.

 From our iMPS groundstates for the Hamiltonian of Eq. (\ref{Ham}),
 we calculate the quantum mutual information for the adjacent two spins.
 In Fig. \ref{fig1}(a), the quantum mutual information $I(R/J)$ is plotted
 as a function of the interaction rate $R/J$ for the spin-$1$ system.
 Straightforwardly, one can notice the discontinuous behaviors of the quantum mutual information
 at the corresponding discontinuous points of the quantum coherence measures.
 As one may expect, this non-analytical behavior is consistent
 with those of all quantum coherence measures.
 The discontinuous behaviors of all tools of quantum information theory indicate
 the occurrence of discontinuous quantum phase transitions at the points $R=\pm R_c$.
 Noticeably, the quantum mutual information vanishes to be zero for $-R_c < R < R_c$.
 This region of zero quantum mutual information corresponds to that of the incoherent phase
 for $-R_c < R < R_c$.
 In the parameter region,
 the vanishing quantum mutual information
 means that the entanglement between the adjacent two spins becomes zero.
 As a result,
 the groundstate is in a product state in the incoherent phase
 for $-R_c < R < R_c$.

 For the spin-$2$ system, Fig. \ref{fig1}(b) shows a monotonous hill-shape of
 the quantum mutual information that has a maximum value at the incoherent point $R/J=0$.
 Interestingly, compared to the incoherent phase of the spin-$1$ system for $-R_c < R < R_c$,
 where the entanglement between the adjacent two spins is zero,
 the incoherent state at $R/J=0$ for the spin-$2$ system
 has a maximum correlation and thus is not a product state.
 Such a maximum value can indicate a quantum phase transition.
 We calculate the derivatives of the quantum mutual information to search for
 a possible non-analyticity of the quantum mutual information.
 However, as shown in Fig. \ref{fig2}(b), similar to the relative entropy of coherence $C_{re}$
 in Fig. \ref{fig2}(a),
 the first-order and the second-order derivatives of the quantum mutual information
 shows a monotonous change across the incoherent point $R/J=0$.
 Accordingly, similar to the relative entropy of coherence $C_{re}$,
 our quantum mutual information $I$ cannot determine clearly whether
 a quantum phase transition occurs at the incoherent point.
 However, it should be noted that although
 the quantum mutual information $I$ measures a different fundamental nature of groundstate
 from the relative entropy of coherence,
 it behaves in accordance with the relative entropy of coherence $C_{re}$
 in detecting quantum phase transition.
 In this aspect, it is possible to say that for the spin-$2$ system,
 the quantum mutual information $I$ and the relative entropy of coherence $C_{re}$
 behave conflictingly to the $l_1$ norm of coherence $C_{l_1}$
 and the quantum Jensen-Shannon divergence $C_{JS}$.
 Then we will study the non-analyticity of groundstate energy to solve this interesting issue
 in the following section.

%
\begin{figure}
\includegraphics [width=0.45\textwidth]{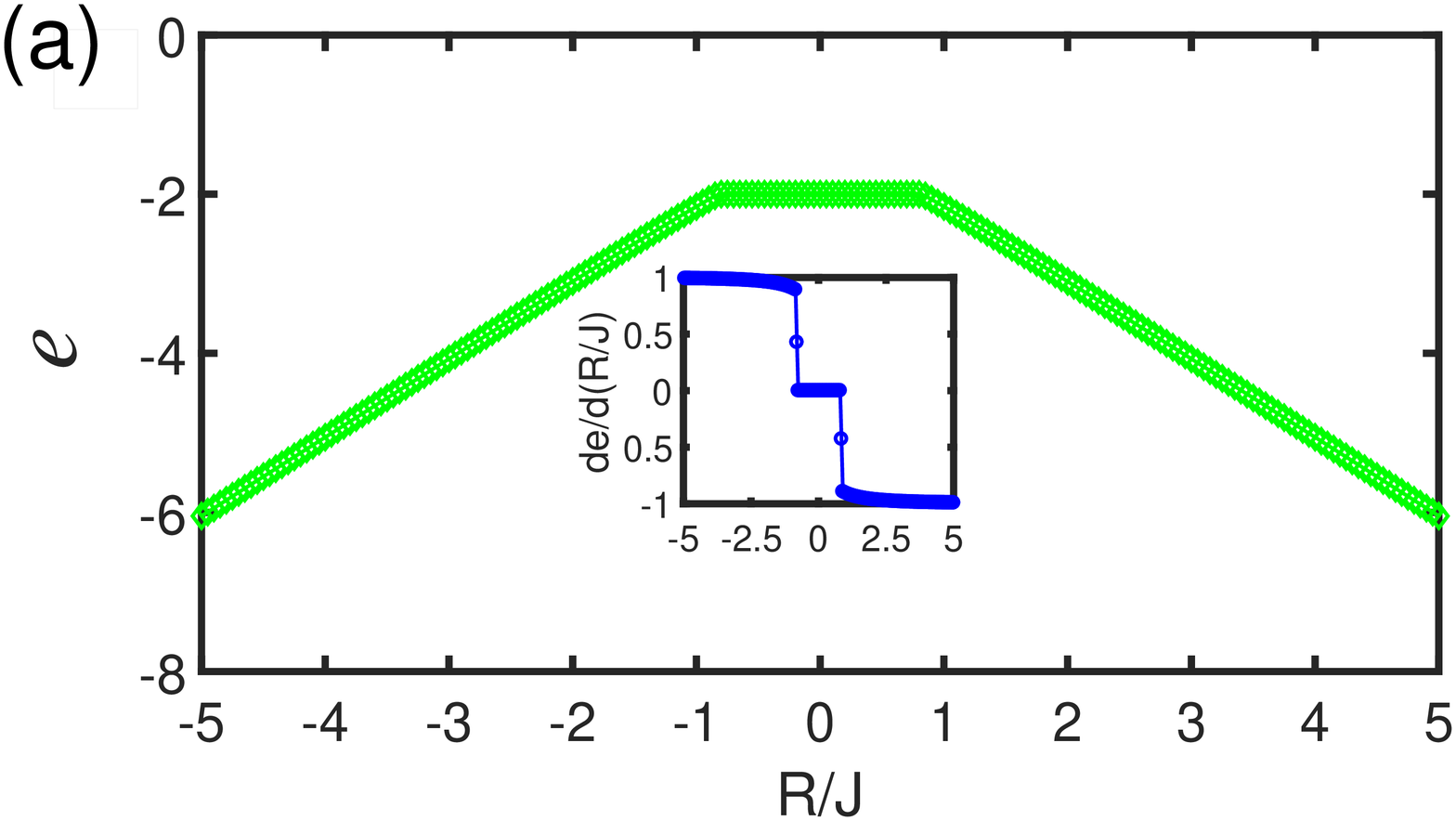}
\includegraphics [width=0.45\textwidth]{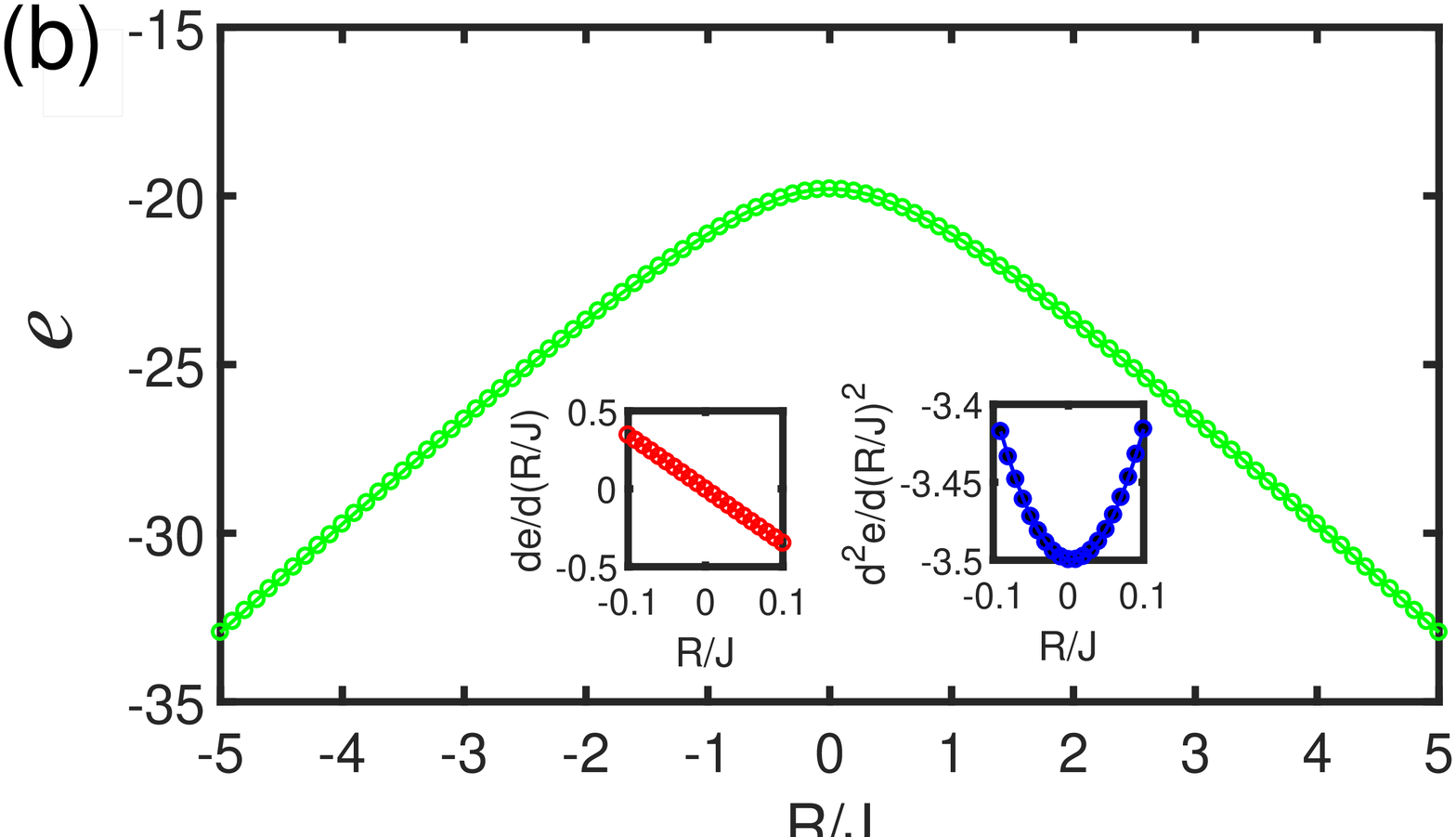}
\caption{ (Color online)
 Groundstate energy per site $e$ as a function of $R/J$ for (a) the spin-$1$ system and (b) the spin-$2$ system.
 In the inset of (a), the first-order derivative of groundstate energy per site is plotted for the spin-$1$ system.
 Also, we plot
 the first- and the second-order derivatives of groundstate energy per site in the inset of (b) for the spin-$2$ system.
}
 \label{fig3}
\end{figure}
\section{Groundstate energy, entanglement entropy, and quantum phase transition}
\label{section4}
 So far, we have discussed about the behaviors of the tools of quantum information theory
 in characterizing quantum phase transitions.
 However, for the spin-$2$ biquadratic XY chain with rhombic single-ion anisotropy,
 the quantum coherence measures detect the conflicting behavior each other
 although the quantum mutual information $I$ is supportive to the relative entropy of coherence $C_{re}$.
 As a standard framework, quantum phase transitions are actually connected with
 the intrinsic features of groundstate energy of the quantum many-body systems, i.e., the energy level crossings which lead to the appearance of non-analyticities of groundstate energy.
 Thus we discuss the behaviors of the groundstate energy per site $e$ to resolve the conflicting issue for the spin-$2$ system in this section.

 From our iMPS groundstates for the Hamiltonian of Eq. (\ref{Ham}),
 one can calculate the groundstate energy. In Fig.\ref{fig3}, the
 groundstate energy per site $e$ is
 plotted as a function of the interaction rate $R/J$ for (a) the spin-$1$ and (b) spin-$2$ systems.
 In the insets, the first- and second-order derivatives of the groundstate energy
 are also plotted.
 In Fig. \ref{fig3}(a) for the spin-$1$ system, the groundstate energy $e(R/J)$
 shows clearly the two sharp kinks
 indicating energy level crossings at $R = \pm R_c$
 and the first-derivative of the groundstate energy also shows its discontinuities
 at the same points, which indicates that first-order quantum phase transitions take place.
 Thus this result manifests that
 the discontinuities of the three coherence measures and the quantum mutual information in Sec. \ref{section3} detect
 the occurrences of the first-order (discontinuous) quantum phase transitions at
 the critical points $R= \pm R_c$.
 Contrastively to the spin-$1$ system,
 Fig. \ref{fig3}(b) shows that
 the groundstate energy is monotonous
 and any noticeable significant change for possible phase transitions
 around $R/J=0$ is not seen in the first- and second-derivatives of the groundstate energy.
 Then
 the monotonous behaviors of the groundstate energy
 could not also solve the conflicting behaviors of the quantum coherence measures, i.e.,
 whether quantum phase transition happens or not at $R/J=0$ for the spin-$2$ system.
 However, it should be noted that the groundstate energy $e$
 behaves in accordance with the quantum mutual information $I$  and the relative entropy of coherence $C_{re}$
 in detecting quantum phase transitions.

\begin{figure}
\includegraphics [width=0.45\textwidth]{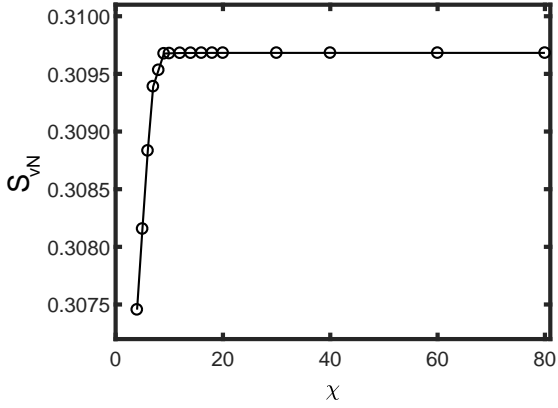}
\caption{ (Color online)
 Bipartite entanglement entropy $S_{vN}(\chi)$ as a function of the truncation dimension $\chi$
 at the incoherent point $R/J=0$ for the spin-$2$ system.
}
 \label{fig4}
\end{figure}

 Actually, our iMPS approach provide a way to detecting critical systems
 by using its characteristic scaling property of bipartite entanglement entropy
 \cite{Su12,Su13,Wang,Dai17,Osterloh,Amico,Korepin,Calabrese,Tagliacozzo}.
 In the iMPS approach, it is known that the bipartite entanglement entropy
 diverges  at a given parameter point in a critical system
 as the truncation dimension $\chi$ increases.
 At critical points for continuous phase transitions,
 the bipartite entanglement entropy also diverges as the truncation dimension $\chi$ increases.
 In order to get more insight into the conflicting point,
 let us then consider the bipartite entanglement entropy for the spin-$2$ system.
 Thus we calculate the bipartite entanglement entropy
 (the von Neumann entropy) by considering the bipartitioned two semi-infinite chains
 in the iMPS representation \cite{Su12,Su13,Wang,Dai17}.
 We plot the bipartite entanglement entropy $S_{vN}(\chi)$ as a function of
 the truncation dimension $\chi$ at $R/J=0$ in Fig. \ref{fig4}.
 Figure \ref{fig4} shows clearly that the entanglement entropy does not diverge but converges
 as the truncation dimension increases.
 Even the entanglement entropy does not change much with the increase of the truncation dimension for higher truncation dimension than $\chi=30$.
 Accordingly,
 this result clarifies that the groundstate is not critical at the incoherent point $R/J=0$
 and then any continuous quantum phase transition across the incoherence point does not take place.
 Since the groundstate energy as well as  the quantum mutual information and the relative entropy
 of coherence is continuous, any possibility occurring a discontinues quantum phase
 can be discarded.
 Conclusively,
 together with the groundstate energy, the mutual information and the relative entropy of coherence,
 the result of the bipartite entanglement entropy represents no occurrence of explicit
 quantum phase transition at $R/J=0$ for the spin-$2$ system.

 Such a conclusion leaves two consequent unconformable facts.
 In the aspect of the quantum coherence measures,
 the non-analytical behaviors of the $l_1$ norm of coherence $C_{l_1}$
 and the quantum Jensen-Shannon divergence $C_{JS}$
 cannot be interpreted as an indication of quantum phase transition at the incoherent point $R/J=0$
 for the spin-$2$ system.
 In the viewpoint of quantum phase and quantum phase transition,
 more intriguing thing is the fact that
 the incoherent point has nothing to do with any continuous
 or discontinuous quantum phase transition.
 This fact that the spin-$2$ system does not undergo any explicit phase transition
 implies that the spin-$2$ system has a very different phase diagram of groundstate from the spin-$1$ system
 even though they have the same form of the Hamiltonian (\ref{Ham}).
 It shows that spin-$2$  has fundamentally different nature from spin-$1$.
 Essential difference between the spin-$1$ and -$2$ systems
 will be clearly clarified by discussing the phases and the order parameters in the next section.

\section{Quantum spin nematic phase transitions}
\label{section5}
 In order to identify the quantum phases in the biquadratic XY chains with rhombic single-ion anisotropy, let us first consider the local magnetizations $\langle S^\alpha_i \rangle$. Existing a local magnetization implies that the individual spins in the chain are oriented in a direction in spin space. Such a spin ordered state is induced by spontaneous breaking both spin-rotation and time-reversal symmetries. Normally, external fields can break such symmetries and can make spins ordered in a direction.
 For the both spin-$1$ and -$2$ biquadratic XY chains with rhombic single-ion anisotropy, however we find that all components of magnetization are zero for the whole parameter space, i.e.,
 $\langle S^\alpha_i \rangle = 0$ for the groundstates,
 which implies that all states of our model preserve time-reversal symmetry for the whole parameter space.
 According to the results of the groundstate energy and the coherence measures,
 thus our systems can have quantum phases without magnetic order, i.e., the so-called spin nematic phases.
 According to the results of the quantum coherence measures, the quantum mutual information,
 and the groundstate energy, quantum phase transitions can occur between spin nematic phases
 in our spin systems.
 Behaviors of quadrupole moments will then characterize quantum spin nematic phases of our spin systems.

 Let us consider the quadrupole moment measured by products of spin operators.
 A symmetric and traceless rank-$2$ quadrupole tensor operator \cite{Penc} can be given by
\begin{equation}
 Q^{\alpha\beta}_i= \frac{1}{2} \left( S^\alpha_i S^\beta_i +  S^\beta_i S^\alpha_i \right) - \frac{1}{3} \mbox{\boldmath \( S \) }^2_i \delta_{\alpha\beta}
\end{equation}
 with $\alpha,\beta \in \{x, y,z\}$ and the Kronecker-delta $\delta_{\alpha\beta}$ at site $i$.
 The actual independent components of quadrupole tensor operator are five
 because $\sum_{\alpha} Q^{\alpha\alpha}_i=0$ for $\alpha = \beta$ and $Q^{\alpha\beta}=Q^{\beta\alpha}$ for $\alpha \neq \beta$.
 We find that all off-diagonal components of quadrupole moments are zero for the whole parameter space, i.e.,
 $\langle Q^{xy}_i \rangle = \langle Q^{yz}_i \rangle=\langle Q^{zx}_i \rangle=0$
 for the both spin-$1$ and -$2$ groundstates.
 Thus two actual components of quadrupole tensor operator can identify quantum phases in our spin systems.
 We calculate the two quadrupole orders given as
\begin{eqnarray}
 Q^{x^2-y^2}_i &=& Q^{xx}_i - Q^{yy}_i,  \\ 
 Q^{3z^2-r^2}_i &=& 3 Q^{zz}_i . 
\end{eqnarray}
 Non-zero quadrupole moments can characterize groundstates breaking spin-rotational symmetry
 by developing an isotropy in their spin fluctuations, which indicates that our spin-$1$ and spin-$2$ systems are in spin nematic quadrupole phases.

\begin{figure}
\includegraphics [width=0.40\textwidth]{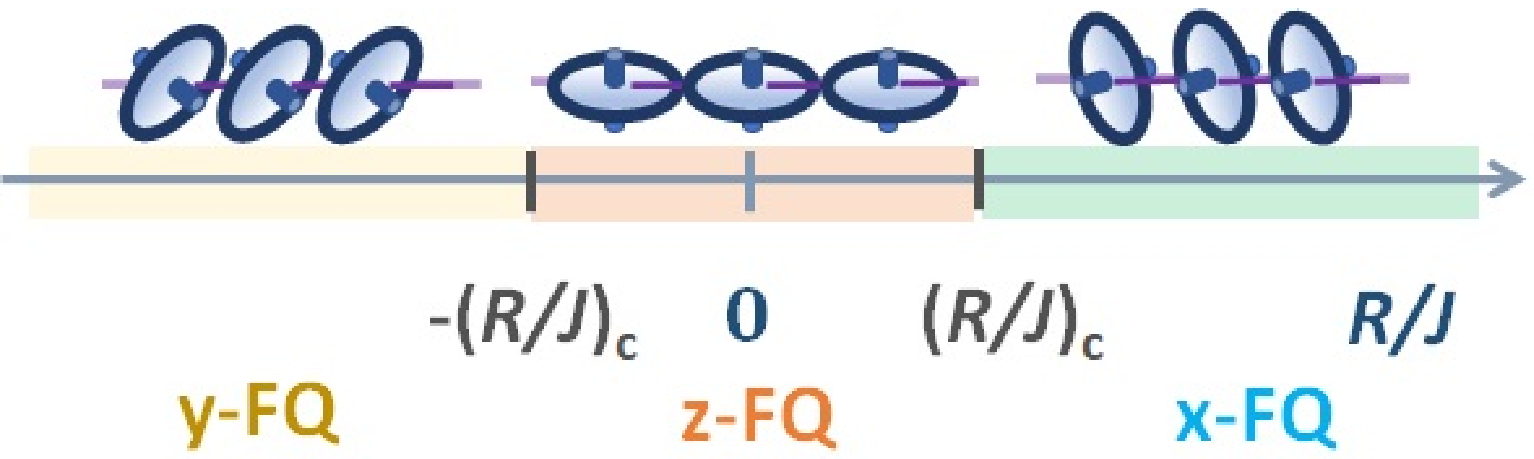}
\includegraphics [width=0.45\textwidth]{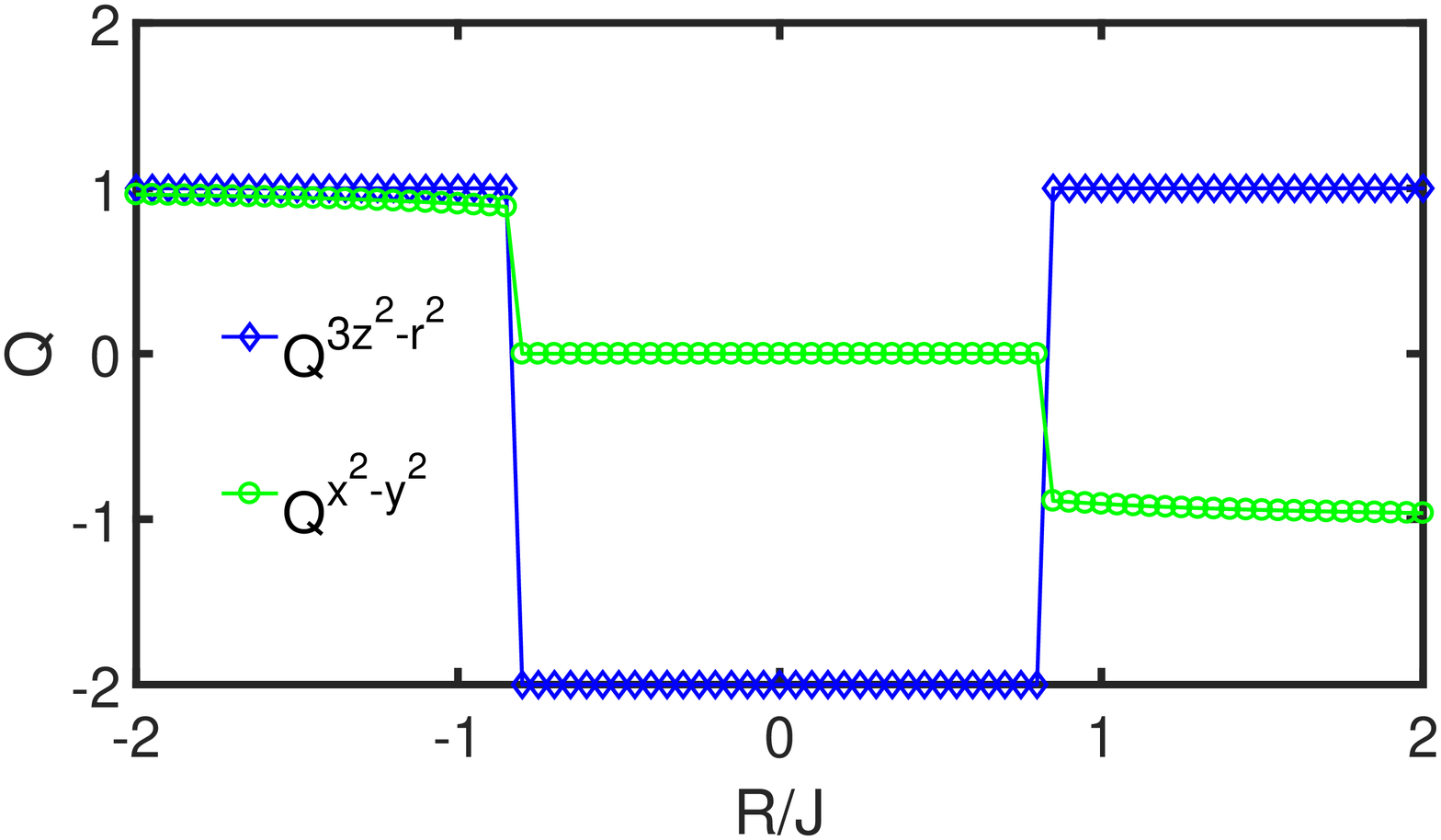}
\caption{Top: Schematic phase diagram.
  Bottom: Quadrupole order parameters $\langle Q^{x^2-y^2} \rangle$ and $\langle Q^{3z^2-r^2} \rangle$  as a function of $R/J$ for the spin-$1$ system.
  In the phase diagram, $\alpha$-$FQ$ denote the ferroquadrupole phases with
  the zero local spin fluctuation
  in the $\alpha$ axis ($\alpha \in \{x,y,z\}$). A disk indicates the local
  spin fluctuation in the plane perpendicular to the $\alpha$ axis at site $i$.}
 \label{fig5}
\end{figure}
\subsection{Quadrupole phases for the spin-$1$ system}
 Let us first discuss about quadrupole phases in the spin-$1$ biquadratic XY chain with rhombic single-ion anisotropy.
 Figure \ref{fig5} displays $\langle Q^{x^2-y^2}_i\rangle$ and $\langle Q^{3z^2-r^2}_i\rangle$ as a function of $R/J$ for the spin-$1$ system, which
 shows clearly the discontinuities of the quandrupole order parameters
 indicating the occurrence of the first-order quantum phase transitions
 as the quantum coherence measures and the quantum mutual information detected.
 In each phase, the quadrupole states can be characterized by the quadrupole moments.
 We identify the phases with the characterization of
 the quadrupole moments as follows.

 (i) The $y$-ferroquadrupole phase with
  $\langle Q^{x^2-y^2}_i\rangle > 0$ for $R < -R_c$.
  As shown in Fig. \ref{fig5}, $\langle Q^{3z^2-r^2}_i\rangle = 1$, i.e.,
  $\langle Q^{zz}_i \rangle = 1/3$ parameter-independently for $R < - R_c$.
  At large negative $R/J$,
   $\langle Q^{x^2-y^2}_i\rangle=\langle Q^{3z^2-r^2}_i\rangle $  gives
  $\langle Q^{xx}_i\rangle = \langle Q^{zz}_i \rangle = -\frac{1}{2}\langle Q^{yy}_i \rangle$, which implies that the groundstate is a uniaxial spin nematic state.
  In the limit of large negative $R/J$, the rhombic single-ion anisotropy becomes predominant
  and thus the local spin state is forced to be the lowest-energy state of
  the anisotropy term $R [ (S^y_i)^2-(S^x_i)^2]$ in the Hamiltonian (\ref{Ham}), which is given by
  $|S^y_i=0\rangle$, where the local spin fluctuates in the $zx$ plane.
  Thus, as shown in the phase diagram in Fig. \ref{fig5},
   with the zero local magnetization $S^\alpha=0$, the local spin fluctuations
  are like a disk in the $zx$ plane and have zero amplitude along the $y$ axis.
  Accordingly, this groundstate is a uniaxial spin nematic state and
  the product state of  $|S^y_i=0\rangle$.
  This phase is referred to the $y$-ferroquadrupole phase.

  As the negative $R/J$ approaches to the transition point $R=-R_c$ from near $R=-R_c$,
  a little local spin fluctuation arises along the $y$ axis and decreases the same amount of  the local spin fluctuation along the $x$ axis without changing along the $z$ axis
  because the local spin fluctuation does not change for the $y$-ferrroquadrupole phase, i.e.,
  $\langle (S^z_i)^2 \rangle =1$.
  The changes of quadrupole moments  reflect to this little local spin fluctuation in the same manner.
  When the negative $R/J$ crosses the transition point,
  the first-order quantum phase transition takes place at $R=-R_c$.

  (ii) The $z$-ferroquadrupole phase with $\langle Q^{x^2-y^2}_i\rangle =0$ for $-R_c < R < R_c$.
  Figure \ref{fig5} shows that $\langle Q^{3z^2-r^2}_i\rangle = -2$, i.e.,
  $\langle Q^{zz}_i \rangle = - 2/3$ and
  $\langle Q^{x^2-y^2}_i\rangle=0$  parameter-independently for $-R_c < R < R_c$.
  This fact gives
  $\langle Q^{xx}_i\rangle = \langle Q^{yy}_i \rangle = -\frac{1}{2}\langle Q^{zz}_i \rangle$. Straightforwardly, the local spin fluctuations are given as
  $ \langle \left( S^{x} \right)^2 \rangle = \langle \left( S^{y} \right)^2 \rangle = 1$ and $ \langle \left( S^{z} \right)^2 \rangle =0$, which means that the groundstate is a uniaxial spin nematic state.
  Thus the local spin fluctuates only in the $xy$ plane, with no change in the $z$ axis
  for the parameter range of the phase.
  As noticed in the quantum mutual information, the groundstate is in a product state, i.e.,
  the product state of  $|S^z_i=0\rangle$ for the parameter range $-R_c < R < R_c$.
  Due to the robust local spin fluctuation in the $xy$ plane for the phase,
  the groundstate structure in the product state does not change until the rhombic single-ion anisotropy overwhelms the biquadratic interaction to induce the sudden change
  of the spin fluctuation at the critical points $R=\pm R_c$.
  This phase is referred to the $z$-ferroquadrupole phase.

  (iii) The $x$-ferroquadrupole phase with $\langle Q^{x^2-y^2}_i\rangle > 0$ for $R  > R_c$.
   This phase has the constant quadrupole order  $\langle Q^{3z^2-r^2}_i\rangle = 1$, i.e.,
  $\langle Q^{zz}_i \rangle = 1/3$.
  At large positive $R/J$,
  as shown in Fig. \ref{fig5}, $\langle Q^{x^2-y^2}_i\rangle=-\langle Q^{3z^2-r^2}_i\rangle = -1$  gives
  $\langle Q^{yy}_i\rangle = \langle Q^{zz}_i \rangle = -\frac{1}{2}\langle Q^{xx}_i \rangle$, indicating that the groundstate is a uniaxial spin nematic state.
  This positive $R/J$ is equivalent to exchanging the $x$ axis and $y$ axis
  of the rhombic single-ion anisotropy term $R [ (S^y_i)^2-(S^x_i)^2]$ in the Hamiltonian (\ref{Ham})
  for the case of the $y$-ferroquadrupole phase.
  The local spin state is forced to be the lowest-energy state of
  the anisotropy term $R [ (S^x_i)^2-(S^y_i)^2]$ in the Hamiltonian, which is given by
  $|S^x_i=0\rangle$, where the local spin fluctuates in the $yz$ plane.
  As a result, the groundstate is a uniaxial spin nematic state and
  the product state of  $|S^x_i=0\rangle$.
  This phase is referred to as the $x$-ferroquadrupole phase.

  Similarly to the $y$-ferroqudrupole phase,
  as the positive $R/J$ approaches to the transition point $R=R_c$ from near $R=R_c$
  in the $x$-ferroquadrupole phase,
  a little local spin fluctuation arises along the $x$ axis and decreases the same amount of  the local spin fluctuation along the $y$ axis without changing along the $z$ axis
  because the local spin fluctuation does not change for the $y$-ferrroquadrupole phase, i.e.,
  $\langle (S^z_i)^2 \rangle =1$..
  When the positive $R/J$ crosses the transition point
  in the $x$-ferroquadrupole phase, the firs-order quantum phase transition occurs at $R=R_c$.

\subsection{Quantum crossover for the spin-$2$ system}
\begin{figure}
\includegraphics [width=0.45\textwidth]{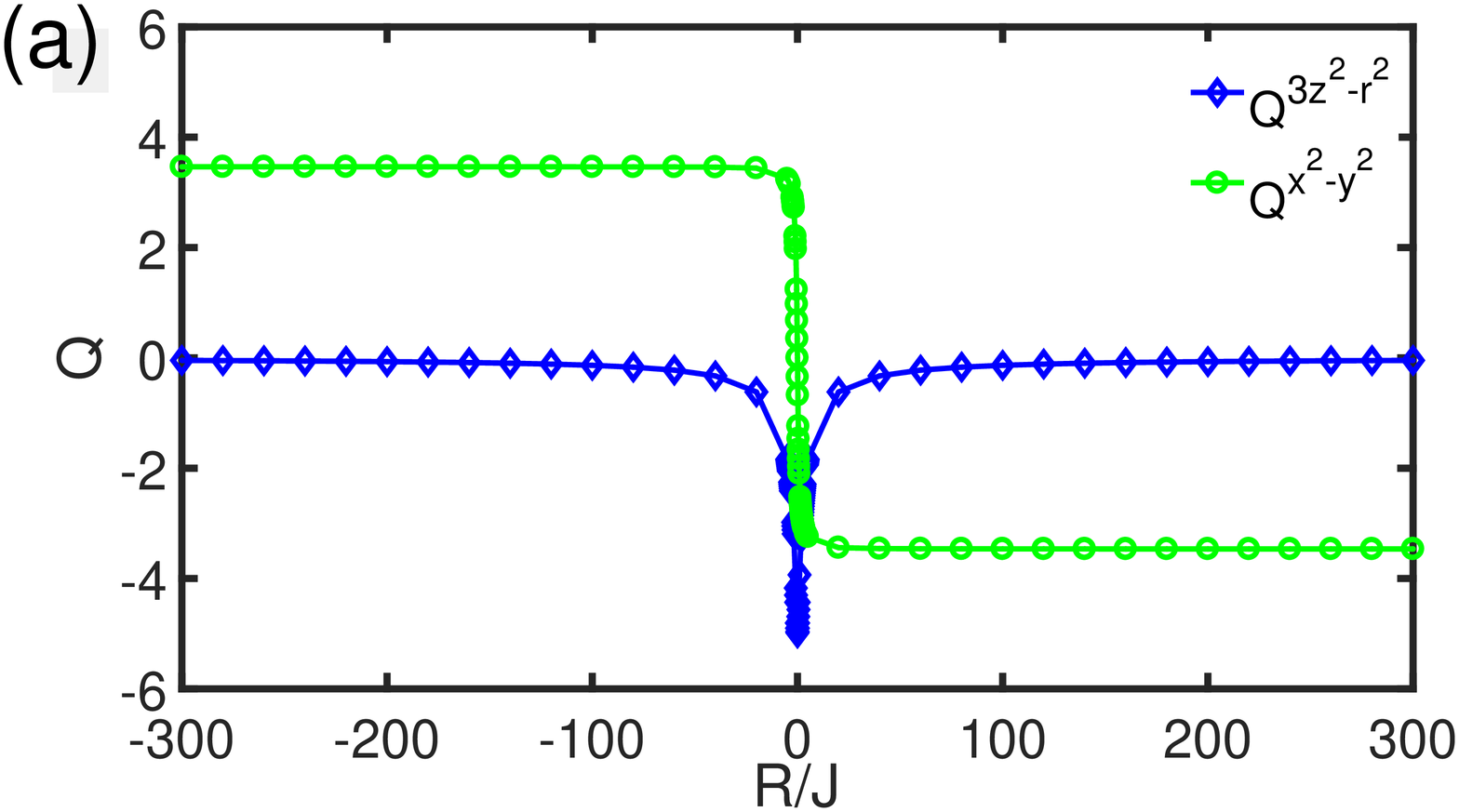}
\includegraphics [width=0.45\textwidth]{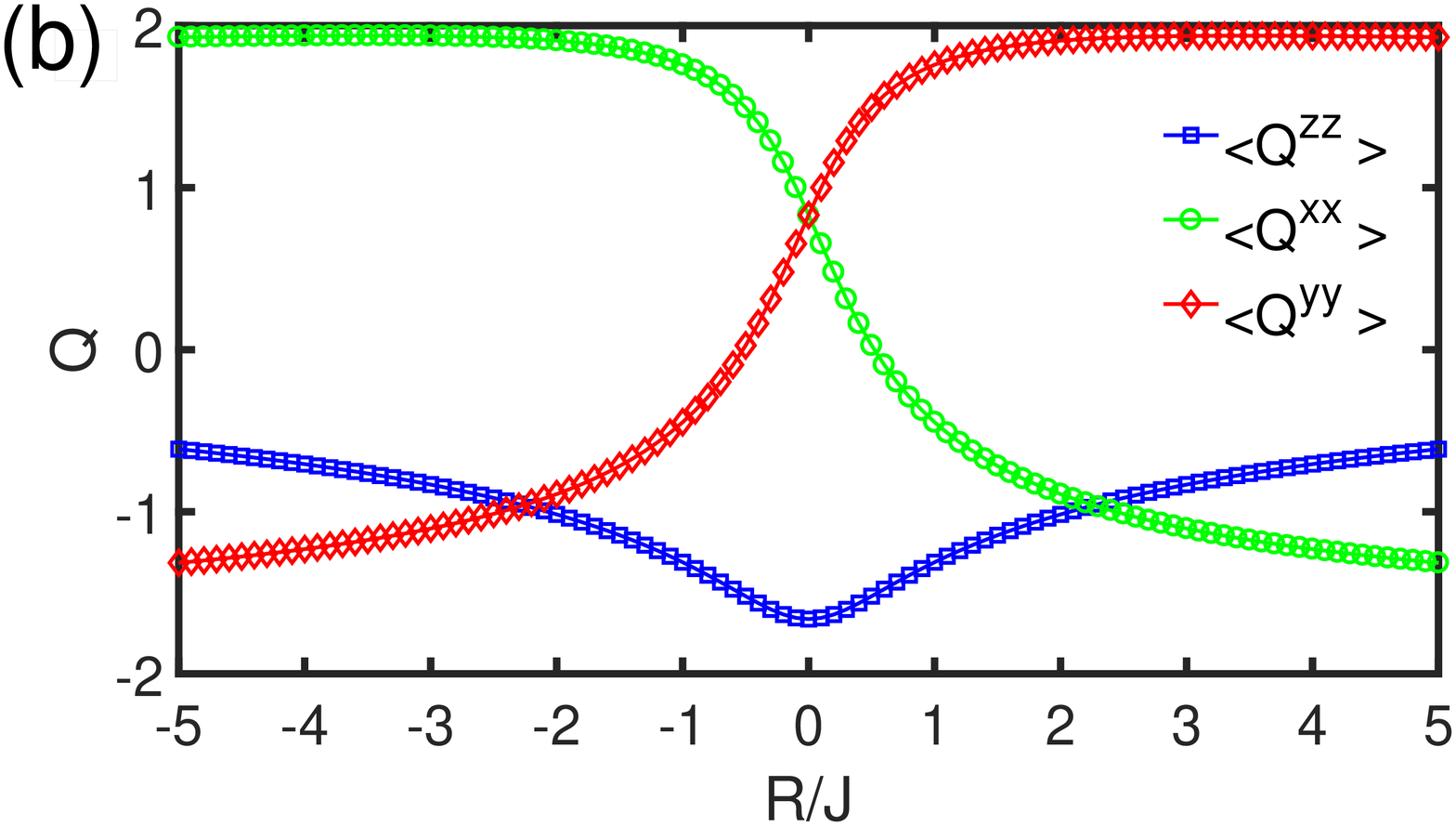}
\caption{
  (a) Quadrupole order parameters $\langle Q^{x^2-y^2} \rangle$ and $\langle Q^{3z^2-r^2} \rangle$ and (b) quadrupole moments $\langle Q^{\alpha\alpha} \rangle$ as a function of $R/J$ for the spin-$2$ system.}
 \label{fig6}
\end{figure}

 Next, we study the spin-$2$ biquadratic XY chain with rhombic single-ion anisotropy.
 As one can easily notice, the spin-$2$ quadrupole order parameter $\langle Q^{x^2-y^2}_i\rangle$ in Fig. \ref{fig6}(a) reveals its very different behavior from the discontinuous features of the spin-$1$ quadrupole order parameter in Fig. \ref{fig5}.
 Only a noticeable change of the spin-$2$ quadrupole order parameter $\langle Q^{x^2-y^2}_i\rangle$
 is its sign change at $R=0$,
 which may indicate the two distinct phases according to its sign.
 Thus let us discuss in details of the quadrupole order parameter to clarify
 what kind of phases can exist and whether a phase transition occurs or not.

 (i) The positive biaxial spin nematic phase with $\langle Q^{x^2-y^2}_i\rangle > 0$ for $R < 0$.
  At large negative $R/J$,
   $\langle Q^{x^2-y^2}_i\rangle=2\sqrt{3}$ and $\langle Q^{3z^2-r^2}_i\rangle =0$  gives
  $\langle Q^{xx}_i\rangle = -\langle Q^{yy}_i \rangle = \sqrt{3}$
  and $\langle Q^{zz}_i \rangle=0$.
  In contrast to the spin-$1$ system, then
  the groundstate is a biaxial spin nematic state.
  Similar to the spin-$1$ system,
  in the limit of large negative $R/J$, the spin-$2$ rhombic single-ion anisotropy becomes predominant and thus the local spin state is forced to be the lowest-energy state of
  the rhombic-anisotropy term $R [ (S^y_i)^2-(S^x_i)^2]$ in the Hamiltonian (\ref{Ham}).
  In sharp contrast to the spin-$1$ system, however the local spin state is not given by $|S^y_i=0\rangle$ where the local spin fluctuates in the $zx$ plane for the spin-$1$ system.
   Whereas for the spin-$2$ system,
   the local spin fluctuates along the all axes and the lowest-energy state gives
   $ \langle ( S^{x}_i )^2 \rangle=2+\sqrt{3}$,
   $\langle ( S^{y}_i)^2 \rangle = 2- \sqrt{3}$, and
   $\langle ( S^{z}_i)^2 \rangle = 2$ with $\langle S^\alpha_i \rangle=0$.
  Accordingly, this groundstate is a biaxial spin nematic state and
  the product state of the lowest-energy state of $R [ (S^y_i)^2-(S^x_i)^2]$
  given by $a|S^x_i=0\rangle + b|S^y_i=0\rangle$
  with $a = (1+\sqrt{3})/\sqrt{6}$ and $b=(1-\sqrt{3})/\sqrt{6}$.
  This phase is then referred to the \textit{positive} biaxial spin nematic phase
  for the positive order parameter $\langle Q^{x^2-y^2}_i\rangle > 0$.

 (ii) The groundstate at $R/J=0$, where  $\langle Q^{x^2-y^2}_i\rangle = 0$.
 $\langle Q^{x^2-y^2}_i\rangle = 0$ gives
 $\langle Q^{xx}_i\rangle = \langle Q^{yy}_i \rangle = -\frac{1}{2}\langle Q^{zz}_i \rangle > 0$ at $R/J=0$, as shown in Fig. \ref{fig6}(a).
 Thus the groundstate is not critical and a product state, but it is a uniaxial spin nematic state.
 However, in contrast to the spin-$1$ system,
 the local spin fluctuation is not confined in the $xy$ plane.
 Actually, this point $R=0$ corresponds to the maximum fluctuation point of
 the local spin along the $z$ axis for the whole parameter space, as shown in Fig. \ref{fig6}(b).

  In fact, there are two more such uniaxial nematic states.
  From Fig. \ref{fig6}(b), one can notice that there are the three uniaxial nematic states
  at $R =0$, $R_{a}$, and $-R_a$ with $R_a = 2.264 J$
  in the whole parameter space.
  At $R = - R_a$ and $R_a$ , the quadrupole moments are given as
  $\langle Q^{yy}_i\rangle = \langle Q^{zz}_i \rangle = -\frac{1}{2}\langle Q^{xx}_i \rangle < 0$ and
  $\langle Q^{zz}_i\rangle = \langle Q^{xx}_i \rangle = -\frac{1}{2}\langle Q^{yy}_i \rangle < 0$, respectively.
 For the three uniaxial spin nematic states,
 as we discussed in the spin-$1$ system, the local spin fluctuation is confined in a plane.

 (iii) The negative biaxial nematic phase with  $\langle Q^{x^2-y^2}_i\rangle < 0$ for $R > 0$.
 Contrary to the large negative $R/J$, at large positive $R/J$,
   $\langle Q^{x^2-y^2}_i\rangle=-2\sqrt{3}$ and $\langle Q^{3z^2-r^2}_i\rangle =0$  gives
  $\langle Q^{xx}_i\rangle = -\langle Q^{yy}_i \rangle = \sqrt{3}$
  and $\langle Q^{zz}_i \rangle=0$, i.e.,
  the groundstate is a biaxial spin nematic state.
  Similar to the large negative $R/J$,
  in the limit of large positive $R/J$, the rhombic single-ion anisotropy becomes predominant
  and thus the local spin state is forced to be the lowest-energy state of
  the rhombic-anisotropy term $R [ (S^x_i)^2-(S^y_i)^2]$ in the Hamiltonian (\ref{Ham}).
  The local spin fluctuates in the all axes and the lowest-energy state gives
   $ \langle ( S^{x}_i)^2 \rangle=2-\sqrt{3}$,
   $\langle ( S^{y}_i)^2 \rangle = 2+\sqrt{3}$, and
   $\langle ( S^{z}_i)^2 \rangle = 2$ with $\langle S^\alpha_i \rangle=0$ as they should be.
  Accordingly, this groundstate is a biaxial spin nematic state and
  the product state of the lowest-energy state $R [ (S^x_i)^2-(S^y_i)^2]$
  given by $b|S^x_i=0\rangle + a|S^y_i=0\rangle$.
  This phase is referred to the \textit{negative} biaxial spin nematic phase
  for the negative order parameter $\langle Q^{x^2-y^2}_i\rangle <0$.

  Note that the positive and negative biaxial spin nematic states are orthogonal each other although $|S^x_i=0\rangle$ and $|S^y_i=0\rangle$ are not orthogonal each other.
  Thus depending on the sign of the quadrupole order parameter $\langle Q^{x^2-y^2}_i\rangle$,
  the two distinct biaxial spin nematic phases can be distinguished.
  However, as noticed by the relative entropy of coherence $C_{re}$ and the quantum mutual information $I$
  in Sec. \ref{section3},
  and the groundstate energy per site $e$ in Sec. \ref{section4},
  the spin-$2$ biquadratic XY chain with rhombic single-ion anisotropy does not undergo any explicit phase transition
  as the $R/J$ varies from $R/J = -\infty$ to $R/J = \infty$.
  Accordingly, the two orthogonal biaxial spin nematic states are connected adiabatically without an explicit abrupt phase transition.

  This adiabatic connection between the two orthogonal biaxial spin nematic states
  can be called \textit{quantum crossover} that
  is a substantial change in the nature of many-body groundstate that occurs over a finite range of the system parameter rather than abruptly at a critical point.
 This can be understood by comparing with the quantum phase transition between
 those two biaxial spin nematic states.
 To do this, let us consider the Hamiltonian in Eq. (\ref{Ham}) for $J=0$, i.e.,
 $H_R = R\sum_{i=-\infty}^{\infty} [ (S^x_i)^2 - (S^y_i)^2]$.
 For $R < 0$, similar to the Hamiltonian in Eq. (\ref{Ham}),
 the groundstate is given by the product state of $a|S^x_i=0\rangle + b|S^y_i=0\rangle$.
 While for $R > 0$, the groundstate should be the product state
 of $b|S^x_i=0\rangle + a|S^y_i=0\rangle$.
 The quandrupole order parameter becomes $\langle Q^{x^2-y^2}_i\rangle=2\sqrt{3}$
 for $R < 0$ and $\langle Q^{x^2-y^2}_i\rangle=-2\sqrt{3}$ for $R > 0$, where
 $\langle Q^{3z^2-r^2}\rangle = 0$.
 It is shown clearly that the groundstate changes abruptly at $R=0$.
 The discontinuity of the quadrupole order parameter indicates
 an occurrence of discontinuous quantum phase transition between
 the positive and negative biaxial spin nematic phases at $R=0$.
 In sharp contrast to this discontinuous quantum phase transition at $R=0$,
 as a consequence,
 the spin-$2$ biquadratic XY chain with rhombic single-ion anisotropy undergoes
 the quantum crossover between the two biaxial spin nematic states.

\begin{figure}
\includegraphics [width=0.45\textwidth]{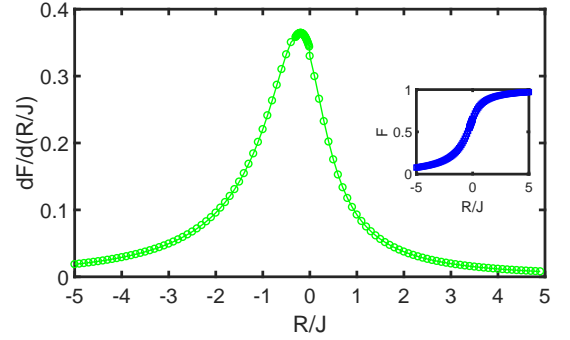}
\caption{ First-order derivative of quantum fidelity per site $d(R/J)$
  as a function of $R/J$ for the spin-$2$ system.
  In the inset, the quantum fidelity is plotted.}
 \label{fig7}
\end{figure}

 The substantial change of groundstate wavefunction structure for the quantum crossover
 can be quantified by directly comparing the groundstate wavefunctions.
 Actually, one can define the overlap function between the groundstate wavefunctions
 as, for instance,
 $f(R/J) =\langle \psi_G(R/J) | \psi_G(+\infty)\rangle$ that ranges $0 \leq f(R/J) \leq 1$
 because the positive and negative biaxial spin nematic states are orthogonal each other,
 i.e., $\langle \psi_G(-\infty) | \psi_G(+\infty)\rangle = 0$, and
 the overlap function of the positive/negative biaxial spin nematic states by themselves
 give  $\langle \psi_G(\pm\infty) | \psi_G(\pm\infty)\rangle = 1$,
 where $|\psi_G(R/J)\rangle$ is the groundstate wavefunction at $R/J$.
 With one of the biaxial spin nematic states as a reference state, the rapidity of the change occurring in the groundstate wavefunction structure is to be the slope of the overlap function, i.e.,
 the derivative of the overlap function with respect to the variable $R/J$.
 Recently, the fidelity susceptibility \cite{Khan} has been used to estimate
 the crossover region in the BCS-BEC crossover.
 Without the loss of generality, we then consider the groundstate quantum fidelity per site $F(R/J)$ \cite{HQZhou,Su13} given as
 $\ln F(R/J) = \lim_{L \rightarrow \infty} \ln f(\langle \psi_G(R/J) | \psi_G(+\infty)\rangle)/L$
 in our iMPS framework.
 Figure \ref{fig7} displays the first-derivative of the quantum fidelity per site $F(R/J)$ as
 a function of $R/J$ with the $F(R/J)$ in the inset.
 The most rapid change of
 the groundstate wavefunction structure takes place with the value $dF/d(R/J) = 0.364$ at $R/J=-0.19$,
 which is located in the middle of a rapid changing some region.
 Then as a strict definition for quantum crossover region,
 one can choose
 the full width at half maximum (FWHM) of the peak in the derivative of quantum fidelity per site, i.e., as an example,
 $-1.21 J \lesssim R_{cross} \lesssim 0.49 J$ from Fig. \ref{fig7}.
 As one can confirm in Figs. \ref{fig6}(a) and (b),
 the quadrupole moments change relatively rapidly for the quantum crossover region estimated from
 the quantum fidelity.
 However, as a less-strict definition for quantum crossover region
 based on the significant changing behavior of the quadrupole moments,
 one can also choose the region in between the two uniaxial spin nematic points, i.e.,
 $ - R_a \lesssim R_{cross} \lesssim R_a$, as can be noticed in Fig. \ref{fig6}(b).

\section{Summary}
\label{summary}

 We have investigated quantum coherence in the groundstate of infinite biquadratic spin-$1$ and -$2$ XY chains with rhombic single-ion anisotropy
 by employing the iMPS representation with the iTEBD method.
 The three quantum coherence measures such as the $l_1$ norm of coherence $C_{l_1}$,
 the relative entropy of coherence $C_{re}$, and the quantum Jensen-Shannon divergence $C_{JS}$,
 and the quantum mutual information have been calculated for the iMPS groundstates.
 In fact,
 the spin-$1$ and -$2$ systems reveal very difference features of quantum phases and phase transitions
 each other although their Hamiltonians have the same form in Eq. (\ref{Ham}).

 For the spin-$1$ system, all of the physical quantities including the groundstate energy, we have considered, have solidly captured the two discontinuous quantum phase transitions
 by means of their non-analytical behaviors at the critical points $R=\pm R_c$.
 However, for the spin-$2$ system,
 the $l_1$ norm of coherence $C_{l_1}$ and the quantum Jensen-Shannon divergence $C_{JS}$
 behaves non-analytically, which may indicate a phase transition similar to the case of
 the spin-$1$ system,
 at the vanishing rhombic-anisotropy point $R=0$.
 Contrary to them, not only the relative entropy of coherence $C_{re}$ and
 the mutual information $I$ but also the groundstate energy do not show any non-analytical
 behavior up to their second-order derivatives.
  However, in our iMPS approach,
  the saturation behavior of the bipartite entanglement entropy
  with the increase of the truncation dimension $\chi$ manifests that
  the spin-$2$ system is not critical at $R=0$, which rules out a possibility of occurring
  a continuous quantum phase transition.
  Accordingly, it was shown that the spin-$2$ system does undergo
  no explicit abrupt phase separation for the whole parameter space.

 In order to determine the phases and the quantum phase transitions in the biquadratic spin-$1$ and -$2$ XY chains with rhombic single-ion anisotropy, the local magnetic moments and quadrupole moments have been investigated.
 We found that for the both spin-$1$ and -$2$ systems,
 the local magnetic moments are zero, i.e., $\langle S^\alpha_i \rangle = 0$ for the whole parameter space.
 For the spin-$1$ system, the local spin quadrupole order parameter $\langle Q^{x^2-y^2}_i \rangle$
 characterize the three distinct uniaxial spin nematic phases by using the three different quadrupole orderings. Explicit discontinuous  quantum phase transitions between
 the three uniaxial spin nematic phases have exhibited in the quadrupole order parameters.

 In contrast to the spin-$1$ system, according to the sign change of the quadrupole order parameter $\langle Q^{x^2-y^2}_i \rangle$, the two biaxial nematic phases can be separated
 at the vanishing rhombic-anisotropy point $R=0$ in
 the biquadratic spin-$2$ XY chain with rhombic single-ion anisotropy.
 As expected from the relative entropy of coherence $C_{re}$ and
 the quantum mutual information $I$,
 no explicit phase transition between the two biaxial spin nematic
 phases was seen in the quadrupole order parameters.
 However, one biaxial spin nematic state is connected to the other biaxial spin nematic state
 by varying the system parameter through the vanishing rhombic-anisotropy point $R=0$.
 Then this phase change can be called the quantum crossover, which was discussed by comparing
 with the explicit discontinuous phase transition in the Hamiltonian with $R=0$.
 Furthermore, the quantum crossover region was estimated by using the quantum fidelity.
 As a result,
 in the sense that the quantum crossover between the two biaxial spin nematic phases
 does not undergo an abrupt change at a critical point,
 the relative entropy of coherence $C_{re}$ and the quantum mutual
 information $I$ exhibit a proper behavior without any abrupt change,
 while the $l_1$ norm of coherence $C_{l_1}$ and
 the quantum Jensen-Shannon divergence $C_{JS}$ disclose
 an ambiguous singular behavior of their first-order derivatives.

\acknowledgments
 Y.-W. Dai acknowledges support in part from the National Natural Science Foundation of China Grant No. 11805285.

\end{document}